\documentclass[osajnl,twocolumn,showpacs]{revtex4}  

\usepackage{graphicx}

\newcommand{\Vec}[1]{\ensuremath{\mathbf{#1}}}
\newcommand{\Sqrt}[1]{\ensuremath{#1^{1/2}}}

\newcommand{\fT}{\ensuremath{\mathrm{fT}}}
\newcommand{\pT}{\ensuremath{\mathrm{pT}}}
\newcommand{\nT}{\ensuremath{\mathrm{nT}}}
\newcommand{\uT}{\ensuremath{\mathrm{\mu T}}}

\newcommand{\T}{\ensuremath{\mathrm{T}}}
\newcommand{\Btot}{\ensuremath{\Vec{B}_{\mathrm{tot}}}}
\newcommand{\Bstat}{\ensuremath{\Vec{B}_{0}}}
\newcommand{\Bosci}{\ensuremath{\Vec{B}_{1}}}
\newcommand{\Dwhw}{\ensuremath{\Delta \omega_\mathrm{HW}}}
\newcommand{\Dwfhw}{\ensuremath{\Delta \omega^\varphi_\mathrm{HW}}}
\newcommand{\Hz}{\ensuremath{\mathrm{Hz}}}
\newcommand{\mus}{\ensuremath{\mu \mathrm{s} }}

\newcommand{\cm}{\ensuremath{\mathrm{cm}}}

\newcommand{\mum}{\ensuremath{\mu \mathrm{m} }}
\newcommand{\degree}{\ensuremath{^\circ}}
\newcommand{\snr}{\ensuremath{R_{\mathrm{SN}}}}
\newcommand{\cramer}{Cram{\'e}r--Rao}
\newcommand{\rf}{_{\mathrm{rf}}}
\newcommand{\pip}{\ensuremath{P_\mathrm{ip}}}
\newcommand{\pqu}{\ensuremath{P_\mathrm{qu}}}
\newcommand{\fbw}{\ensuremath{f_\mathrm{bw}}}
\newcommand{\Fbw}{\ensuremath{F_\mathrm{bw}}}

\begin{document}

\title{Optimization and performance of an optical cardio-magnetometer}

\author{Georg Bison}
 \email{georg.bison@unifr.ch}
\author{Robert Wynands}

 \altaffiliation[Present address: ]
 {Physikalisch--Technische Bundesanstalt,
  Bundes\/allee 100, 38116 Braunschweig, Germany}
\author{Antoine Weis}
 \affiliation{Department of Physics, University of Fribourg,
  Chemin du Mus\'ee 3, 1700 Fribourg, Switzerland}

\date{\today}

\begin{abstract}
Cardiomagnetometry is a growing field of noninvasive medical
diagnostics that has triggered a need for affordable
high-sensitivity magnetometers.
Optical pumping magnetometers are promising candidates satisfying
that need since it was demonstrated that thy can map the heart
magnetic field.
For the optimization of such devices theoretical limits on the
performance as well as an experimental approach is presented.
The promising result is a intrinsic magnetometric sensitivity of
63 fT$/\sqrt{\mathrm{Hz}}$ a measurement bandwidth of 140 Hz and a
spatial resolution of 28 mm.
\end{abstract}


\maketitle

\section{Introduction}

Biomagnetometry is a rapidly growing field of noninvasive medical
diagnostics \cite{andrae}.
In particular, the magnetic fields generated by the human heart
and brain carry valuable information about the underlying
electrophysiological processes \cite{wikswo}.
Since the 1970s superconducting quantum interference devices
(SQUIDs) have been used to detect these generally very weak
biomagnetic fields \cite{cohen}.
The magnetic field of the human heart is the strongest biomagnetic
signal, with a peak amplitude of 100~\pT, but since this is still
orders of magnitude weaker than typical stray field interference the
measurement of such signals could initially only be performed inside
expensive magnetically-shielded rooms (MSR).
Progress in medical research in the past decade has motivated a
need for more affordable cardiomagnetic sensors.
Recently, multichannel SQUIDs were developed that no longer
require shielding due to the use of gradiometric configurations.
Such devices are commercially available but are still quite
expensive in both capital and operational costs.

Optical pumping magnetometers (OPM) have been widely known since
the 1960s \cite{bloom}, and offer both high sensitivity and
reliable operation for research \cite{cohen-tann} and applications
like geomagnetometry \cite{alexandrov3decades}.
Since OPMs usually work with a near room-temperature thermal
alkali metal vapor, they avoid the need for the cryogenic cooling
that makes SQUIDs so costly and maintenance intensive.
Our goal was to develop an affordable, maintenance-free device that
is both sensitive and fast enough to measure the magnetic field of
the human heart.
In order to be competitive with the well-established SQUIDs, a
cardiomagnetic sensor has to offer a magnetic field sensitivity of at
least 1~\pT\ with a bandwidth of about 100~\Hz.
Furthermore, the spatial resolution of the sensor has to be better
than 4~\cm, the standard separation of grid points during mapping.

Since the cardiomagnetometry community is mainly interested in one
of the components of the magnetic field vector, one might think of
using vector-type OPMs like the Hanle magnetometer
\cite{kastlerhe} or the Faraday magnetometer, devices which
operate in zero fields only \cite{budker_pra_2000}.
However, these devices lose their sensitivity in the presence of
even tiny field components in directions perpendicular to the
field of interest.
The broadening caused by such transverse field components must be
kept well below the width of the magnetometer resonance
\cite{budker_pra_2000}, thus limiting those components to values
below a few tenths of~\pT.
Accordingly, optical vector magnetometers cannot be used for
cardiomagnetometry in a straightforward way since the heart field
features time-varying transverse field components on the order of
100~\pT.
We have therefore concentrated on the $M_x$~OPM, which exhibits a
fast response and which has been shown to be sufficiently sensitive in
an unshielded environment.
Furthermore, lamp-pumped $M_x$~OPMs were used for the first
biomagnetic measurements \cite{kozlov1} with optical magnetometers in
the early 1980s, although that work was discontinued.
Instead of lamps, we use diode lasers as a~light source in order to
build a device that will scale to the many channels needed for fast
mapping of the cardiomagnetic field.

\section{Principle of scalar OPM operation}

Optically pumped magnetometers operate on the principle that the
optical properties of a suitable atomic medium are coupled to its
magnetic properties via the atomic spin.
The ensemble average of the magnetic moments associated with the
spins can be treated as a classical magnetization vector $\Vec{M}
= N_F g_F \,\mu_B \langle \Vec{F} \rangle /\hbar$ in space.
Here $N_F \langle \Vec{F} \rangle = N_F\, \mathrm{tr} (\rho_F
\Vec{F})$ with $\mathrm{tr}\, \rho_F =1 $ is the total angular
momentum of $N_F$ atoms in an optical hyperfine level $F$ where
$\rho_F$ is the density matrix and $g_F$ the Land{\'e} factor of the
state.
Optical magnetometers detect changes of the medium's optical
properties induced by the precession of $\Vec{M}$ in a magnetic
field \Btot.
The frequency of this precession, the Larmor frequency $\omega_L$,
is proportional to the modulus of \Btot:
\begin {equation}
 \omega_L
 = \frac{g_F \mu_B}{\hbar} \left| \Btot \right|
 \quad \equiv  \quad  \gamma_F \left| \Btot \right|.
\label{eq:gf}
\end{equation}
%
%
For Cs the constant of proportionality, $\gamma_F$, has a value of
$ 2\pi \times 3.5\: \Hz/\nT$.
All atomic vapor magnetometers measure the magnetic field via a
direct or indirect measurement of the Larmor frequency.

\subsection{The $M_x$ magnetometer}
\begin{figure}
\centerline{\includegraphics{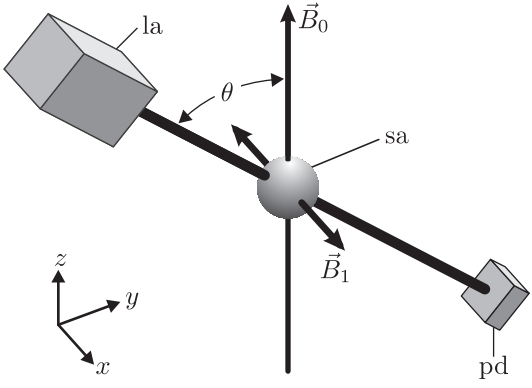}} \caption{ Basic
geometry of the $M_x$-magnetometer setup: The laser (la) emits a
beam that traverses the sample (sa) at angle $\theta$ with respect
to the magnetic field \Bstat.
The transmitted power is detected by a photodetector (pd).
The static magnetic field \Bstat{} is aligned along the
$z$-direction.
The oscillating magnetic field \Bosci{} is aligned along the
$x$-direction. } \label{fig:setup3d}
\end{figure}

In the case of the $M_x$ magnetometer, a magnetic-resonance
technique is used to measure the Larmor frequency directly,
by employing two perpendicular magnetic fields \Bstat{} and \Bosci.
The static magnetic field \Bstat{} is aligned along the
$z$-direction.
As Fig.~\ref{fig:setup3d} shows, the $\Vec{k}$-vector of the laser
beam lies in the $yz$-plane and is oriented at an angle $\theta$
with respect to the $z$-direction.
The magnetometer is sensitive to the modulus of \Bstat.
The oscillating magnetic field $\Bosci=\hat{x} B\rf \cos \omega\rf
t$ is aligned along the $x$-direction with an amplitude much
smaller than $B_0$.

In order to introduce the basic concepts we discuss the simplest case
of an $F=1/2$ state. The motion of $\Vec{M}$ under the influence of
\Bstat{} and \Bosci{} is then given by the Bloch equations:
%
\begin{eqnarray}
  \left( \begin{array}{c} \dot{M}_x \\
                          \dot{M}_y \\
                          \dot{M}_z
           \end{array}  \right)
  &=&
  \left(\begin{array}{c} M_x \\
                         M_y \\
                         M_z
          \end{array}  \right)
  \times
  \left( \begin{array}{c} \gamma_F B\rf 2 \cos\omega\rf t \\
                          0 \\
                          \gamma_F B_0
         \end{array}  \right)
\label{eq:bewgl} \\
  &&{}-\left(\begin{array}{c} \gamma_2 M_x \\
                              \gamma_2 M_y \\
                              \gamma_1 M_z
               \end{array}  \right)
   + \Gamma_P \left( \begin{array}{c}
              \phantom{- M_0 \sin\theta} - M_x \\
              \phantom{-} M_0 \sin\theta - M_y \\
              - M_0 \cos\theta - M_z
   \end{array} \right)\:. \nonumber
\end{eqnarray}
The first term describes the precession of $\Vec{M}$ around the
magnetic fields.
The second term describes the longitudinal ($\gamma_1$) and
transverse ($\gamma_2$) relaxation of $\Vec{M}$.
The third term represents the effect of optical pumping with
circularly polarized light that creates the magnetization.
It can be treated as an additional relaxation leading to an
equilibrium orientation aligned with the $\Vec{k}$-vector of the
incoming light at the pumping rate $\Gamma_P$.
Both relaxations add up to the effective relaxation rates
$\Gamma_{1,2}=\gamma_{1,2}+\Gamma_P$.
In the case of small $B\rf$ amplitudes, Eq.~(\ref{eq:bewgl}) can
be solved using the rotating--wave approximation \cite{millihertz}
which leads to a steady-state solution where $\Vec{M}$ rotates
around \Bstat{} at the driving frequency $\omega\rf$.

The optical property used in the $M_x$ magnetometer is the optical
absorption coefficient which determines the power, $P$, of the light
transmitted through the medium.
For circularly polarized light, the transmitted power is proportional
to the projection of \Vec{M} on the \Vec{k}-vector of the incoming
light.
Therefore, the precessing magnetization results in a modulation of
the absorption index measurable as an oscillation of $P$.
The in-phase and quadrature components of $P$ with respect to the
driving field can be obtained from Eq.~(\ref{eq:bewgl}):
\begin{eqnarray}
 \pip(\delta)& = & -P_0 \sin(2\theta) \frac{\Omega\rf \delta}
             {\Omega\rf^2 \Gamma_2/\Gamma_1 + \Gamma_2^2 +
             \delta^2}\:\mbox{, and}
\label{eq:pip}
 \\
 \pqu(\delta)& = & - P_0 \sin(2\theta) \frac{\Omega\rf \Gamma_2}
             {\Omega\rf^2 \Gamma_2/\Gamma_1 + \Gamma_2^2 +\delta^2}\:.
\label{eq:pqu}
\end{eqnarray}
Here $\Omega\rf= \gamma_F B\rf$ is the Rabi frequency and
$\delta=\omega\rf-\omega_L$ the detuning of the oscillating field
\Bosci{} from the Larmor frequency.
The constant $P_0$ combines all factors such as the initial light
power, the number of atoms in the sample, and the cross section for
light-atom interactions determining the absolute amplitude of the
signal.
The components can be measured using phase-sensitive detection.
The signals are strongest for $\theta = 45^\circ$, which was used in
all experiments.

\begin{figure}
 \centerline{\includegraphics{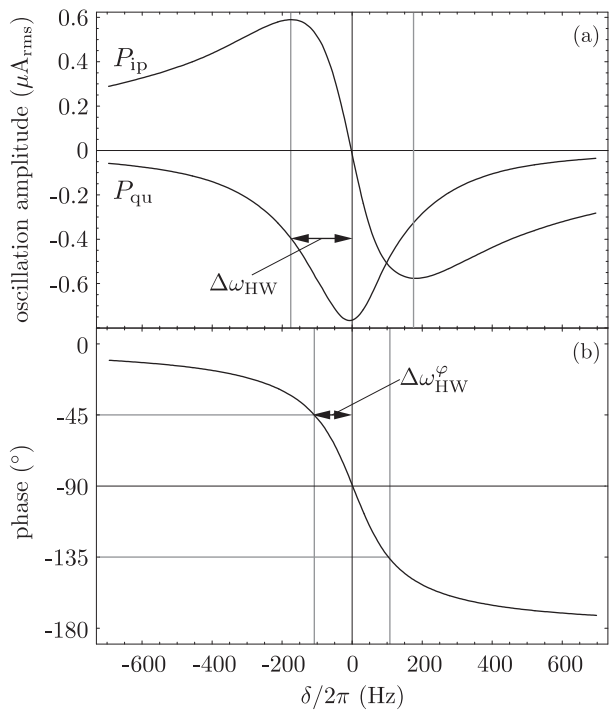}}
 \caption{
(a) Measured magnetic resonance lineshapes of the in-phase ($a_1$)
and quadrature signals ($a_2$), measured in a single sweep of 20~s
with the cardiomagnetometer placed in a poorly shielded room.
Magnetic 50~\Hz{} line interference was suppressed using a
$4^{th}$ order lowpass filter (time constant 10~ms).
The half-width, derived from a fit, was $\Delta\omega_{\mathrm
{HW}}/2 \pi=173\:\Hz$.
(b) Magnetic resonance line shape of the oscillation phase
measured with respect to the driving field $B_{\mathrm {rf}}$.
The data was obtained in real time using a digital lock-in
amplifier.
The fitted half-width is: $\Delta\omega^\varphi_{\mathrm {HW}}/2
\pi=109\:\Hz$. } \label{fig:res}
\end{figure}

Both \pip\ and \pqu\ show resonant behavior near $\delta=0$.
\pqu\ has an absorptive Lorentzian line shape, and \pip\ has a
dispersive Lorentzian line shape with the same half width
expressed as
\begin{equation}
  \Delta\omega_{\mathrm{HW}}=\sqrt{\Omega\rf^2 \Gamma_2/\Gamma_1 + \Gamma_2^2}
                            =\Gamma_2 \sqrt{S+1}\,.
\label{eq:deltaomega}
\end{equation}
Here $S=\Omega^2/(\Gamma_1 \Gamma_2)$ is the saturation parameter
of the rf field.
Figure~\ref{fig:res}(a) shows measured line shapes under conditions
optimized for maximal magnetometric sensitivity (see
Sec.~\ref{sec:optimization} for details).

Signal \pip\ is of particular interest because it has a dispersive
shape, featuring a steep linear zero-crossing at $\delta=0$.
In this region \pip\ can be used to measure the deviation of
\Bstat\ from the value that corresponds to $\omega\rf$.
The same is true for the deviation of the phase difference $\varphi$
between the measured oscillation and the driving field from
-90\degree\ (see Fig.~\ref{fig:res}(b)).
The phase difference $\varphi$ can be calculated from \pip\ and \pqu,
yielding
\begin{equation}
\tan  \varphi =  \frac{\pqu}{\pip} = \frac{\Gamma_2}{\delta}\:.
\end{equation}
The phase signal changes from $\varphi=0$ at low frequencies to
$\varphi=-\pi$ at high frequencies.
For practical reasons it is preferable to shift the phase by
90\degree\ so that it passes through zero in the center of the
resonance ($\delta=0$).
This can easily be done by shifting the reference signal by $\pi/2$
using the corresponding feature of the phase detector.
In mathematical terms that 90\degree\ shift is equivalent to the
transformation $\pqu\rightarrow -\pip$ and $\pip\rightarrow \pqu$,
yielding
\begin{equation}
\tan  \varphi' =  - \frac{\delta}{\Gamma_2}.
\label{eq:phase}
\end{equation}
The width of the phase signal \Dwfhw\ is smaller than \Dwhw\
because it is not affected by rf power broadening, i.e., it is
independent of $\Omega\rf$:
\begin{equation}
\Dwfhw=\Gamma_2<\Dwhw\:.
\end{equation}
The narrower lineshape of the phase signal is exactly compensated
by a better S/N ratio of \pip\, [see Eqs.~(\ref{eq:deltaomega}),
(\ref{eq:dr}), and (\ref{eq:di})] resulting in a statistically
equivalent magnetic field resolution for both signals.
However, since the lineshape of the phase signal depends only on
$\Gamma_2$, it is easier to calibrate in absolute field units.
Furthermore, light amplitude noise, for instance caused by fluctuating
laser intensities, does not directly affect the phase signal, since
both \pip\ and \pqu\ scale in the same way with light intensity.
Only the much weaker coupling via the light shift can cause the phase
signal to reflect light amplitude noise.
Considering those practical advantages of the phase signal we
concentrate in the following sections on the sensitivity of the phase
signal to magnetic field changes.

\subsection{Nyquist plots}
\label{sec:nyquist}

\begin{figure}
\centerline{\includegraphics{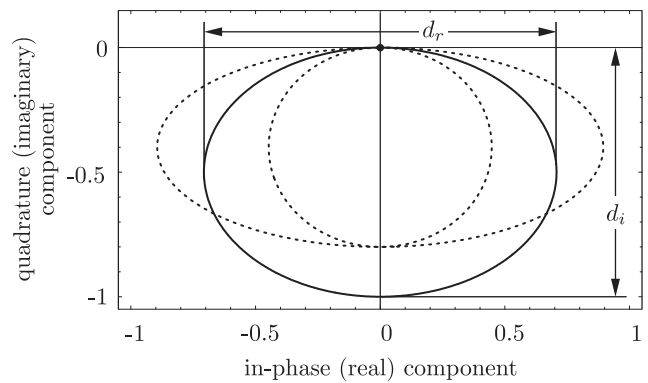}} \caption{
Nyquist plots for three different values of $S$.
The solid line is for $S=1$, the dashed lines are for $S=0.25$
(nearly round) and $S=4$ (elliptical), respectively.
If $\omega\rf$ is scanned towards increasing values the system
evolves clockwise through the Nyquist plot. } \label{fig:circrf}
\end{figure}

The lineshapes $\pip(\delta)$, and $\pqu(\delta)$ of the magnetic
resonance have a major influence on the magnetometric sensitivity.
The magnetic resonances in Eqs.~(\ref{eq:pip}) and~(\ref{eq:pqu}) can
be interpreted as a complex transfer function $t(\omega)$ connecting
the current that drives the rf-coils, $I_i=I_0\exp(i \omega t)+c.c.$,
and the photocurrent $I_p=t(\omega) I_i$ of the photodiode.
By setting the effective transverse relaxation rate $\Gamma_2$ as the
unit of frequency and using the normalized detuning
$x\;(=\delta/\Gamma_2)$, $t$ can be written in dimensionless units as:
\begin{equation}
t=t_0\frac{\sqrt{S} (i+x)}{1+S+x^2}\;.
\label{eq:t}
\end{equation}
A parametric plot of $t(x)$ in the complex plane --- called a
Nyquist plot~--- was found to be useful for the inspection of
experimental data.
In this representation $t(x)$ appears as an ellipse with diameters
$d_r$ and $d_i$ for the real (in-phase) and imaginary (quadrature)
components respectively (see Fig.~\ref{fig:circrf}):
\begin{eqnarray}
d_r &=& t_0\sqrt{\frac{S}{1+S}}
\label{eq:dr} \\
d_i &=& t_0\frac{\sqrt{S}}{1+S}\:.
\label{eq:di}
\end{eqnarray}
The saturation parameter of the rf transition, $S$, can be extracted
from the ratio of the two diameters:
\begin{equation}
S = \frac{\Omega^2}{\Gamma_1 \Gamma_2} =\frac{d_r^2}{d_i^2}-1.
\end{equation}

\begin{figure}
\centerline{\includegraphics{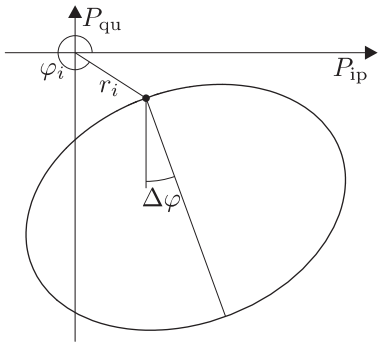}} \caption{
Nyquist plot of a resonance with $S=1$ when an interfering sine
wave of amplitude $r_i$ and phase $\varphi_i$ is added.
A phase offset of $\Delta \varphi$ in the demodulation due to a
poorly adjusted lock-in phase leads to a rotated ellipse. }
\label{fig:circ_phase}
\end{figure}

Figure~\ref{fig:circ_phase} shows a Nyquist plot of a resonance for a
situation in which an interfering sine wave is added to the
photocurrent, leading to a shifted ellipse.
The amplitude $r_i$ and the phase $\varphi_i$ of the interference can
be easily extracted from the Nyquist plot.
A phase shift in the demodulation leads to a rotated ellipse.
In this situation the spectra of in-phase $\pip(\delta)$ and
quadrature $\pqu(\delta)$ components as a function of rf detuning
appear asymmetric.

By means of Nyquist plots it is easy to distinguish between an
asymmetry caused by improper adjustment of the lock-in phase and
one caused by inhomogeneous broadening.
The latter causes a deviation from the elliptical shape.
One model for inhomogeneous broadening is to assume a gradient in the
static magnetic field.
Since we use buffer-gas cells the atoms do not move over large
distances during their spin coherence lifetime so that inhomogeneous
magnetic fields are not averaged out.
Instead, atoms at different locations in the cell see different
magnetic fields, resulting in an inhomogeneous broadening of the
magnetic resonance line.

\begin{figure}
\centerline{\includegraphics{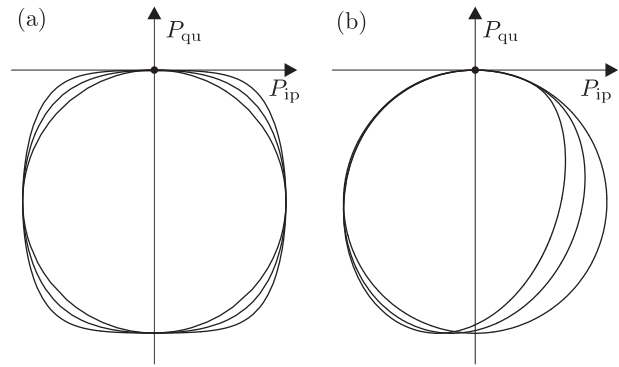}} \caption{
Nyquist plots for different magnetic field distributions, each
scaled to fit in a square of length~1.
Part (a) shows the deviation from circular for constant magnetic
field distributions.
The innermost trace is for an unperturbed resonance.
The two outer traces are calculated for field distribution widths
of $x_g=40\, \Gamma_2$ and $400\, \Gamma_2$, respectively.
Part (b) shows the deviation from circular for linear field
distributions.
The (outer) circular trace is for an unperturbed resonance.
The other two are calculated for distribution widths of $x_g= 5 \,
\Gamma_2$ and $10 \, \Gamma_2$, respectively. }
\label{fig:circ_grads}
\end{figure}

Figure~\ref{fig:circ_grads} shows calculated Nyquist plots for
different gradients of the static field \Bstat.
The simplest model for such an inhomogeneity is a constant gradient
$\mathrm{d} B_z/\mathrm{d}z$ over the length~$l_z$ of the cell.
This is expressed by a convolution of the theoretical magnetic
resonance signals $t(x)$ [see Eq.~(\ref{eq:t})] with the normalized
distribution of magnetic fields $g(x)$ which, in this case, is a
constant over the interval
\begin{equation}
2 x_g = \frac{\gamma_F}{\Gamma_2} l_z dB_z/dz\,.
\end{equation}
Since $g(x)$ vanishes everywhere except for $-x_g\leq x \leq x_g$ the
convoluted resonance $t'$ is given by
\begin{eqnarray}
 t'(x)&=&
 \int_{-{x_g}}^{x_g} \!t(x-x')\, g(x')
 \,\mathrm{d}x'\, ,
\label{eq:convolve}
\end{eqnarray}
which can be evaluated analytically
\begin{eqnarray}
 t' &=&
   \frac{\sqrt{S}}{4\,x_g} \left[\ln \left\{1+S+(x_g - x)^2\right\}\right.
\nonumber \\
   & & \left. - \ln \left\{1+S+(x_g + x)^2 \right\}\right]
\nonumber \\
 &&- \,  \frac{i}{2\,x_g} \sqrt{\frac{S}{1+S}}
         \left\{ \arctan \left(\frac{x_g-x}{\sqrt{1+S}} \right) \right.
\nonumber  \\
 && + \left.  \arctan \left(\frac{x_g+x}{\sqrt{1+S}} \right)
   \right\}\, .
\label{eq:tgrad}
\end{eqnarray}
The main effect of the constant magnetic field distribution is to
broaden the resonance, to decrease the amplitude, and to make the
line shape differ from a Lorentzian.
In the Nyquist plot this is seen by a deformation of the elliptical
trace towards a rectangular trace as shown in
Fig.~\ref{fig:circ_grads}(a).
The effect is clearly visible in Fig.~\ref{fig:circ_grads}(a) for
rather large widths of the magnetic field distribution; in the
experiment, however, the effect can be detected for much smaller
inhomogeneities due to the large signal/noise ratio.

%
%
%
%
%

\section{Experimental Setup}

\begin{figure}
\centerline{\includegraphics{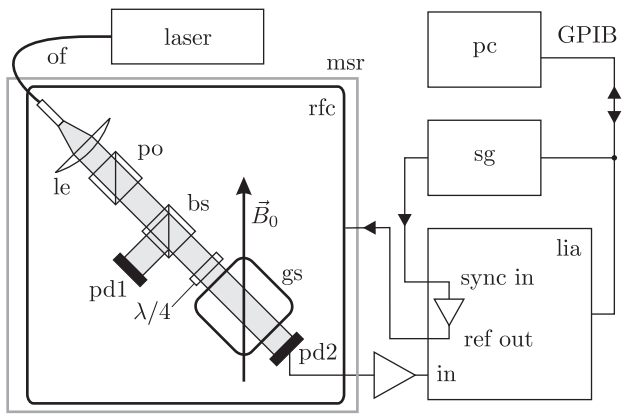}} \caption{
Schematic of the experimental setup.
Light from the diode laser is delivered via an optical fiber (of)
to the experiment in the magnetically shielded room (msr).
The light is collimated by a lens (le) and polarized by a
polarizing beam splitter cube (po).
The beam splitter (bs) reflects 50\% of the beam to photodiode 1
(pd1) used for monitoring of the initial light power.
The remaining beam passes a quarter-wave plate ($\lambda/4$)
providing circular polarized light to the glass cell (gs) that
contains the atomic medium.
Photodiode 2 (pd2) measures the transmitted light intensity.
Its signal is amplified by a current amplifier and fed to the
lock-in amplifier (lia).
The reference output of the lia drives the radio frequency coils
(rfc).
The reference frequency of the lia is controlled by a sweep
generator (sg).
Automatic control and data aquisition is done by a PC via the GPIB
bus. } \label{fig:setup}
\end{figure}

The magnetometer described here was part of the device used by us
to measure the magnetic field of the human heart
\cite{bison1,bison2}.
The setup was designed so that a volunteer could be placed under
the sensor, with his heart close to the glass cell containing
the Cs sample.
For moving the volunteer with respect to the sensor---necessary
for mapping the heart magnetic field---a bed on a low friction
support was used.

The magnetometer sensor head  itself was placed in a room with
moderate magnetic shielding.
The room was $1.7\times 2.3\times 2.5\:\mathrm{m}^3$ in volume
shielded by a 1~mm $\mu$-metal layer and an 8~mm copper-coated
aluminum layer.
For low frequencies, the shielding factor was as low as 5 to 10,
whereas 50~\Hz{} interference was suppressed by a factor of 150.
Inside the shielded room, surrounding the sensor itself, three coil
pairs were placed for the three dimensional control of the magnetic
field.
In the $z$-direction (vertical) two round 1~m diameter coils were
used.
To make room for the patient, the spacing between the coils had to be
62~cm, far away from the Helmholtz optimum of 50~cm.
The two coil pairs for the transverse magnetic fields ($x$~and
$y$~directions) formed four of the faces of a cube 62~cm on a side.
All six coils were driven independently by current sources so that the
sum and the difference of the currents in each coil pair could be
chosen independently.
This allowed us to control not only the magnetic field amplitudes in
all three directions, but also the gradients $\mathrm{d}
B_i/\mathrm{d} i$.
The field components and gradients were adjusted to produce a
homogeneous field of 5~\uT{} in the $z$ direction.

\begin{figure}
\centerline{\includegraphics{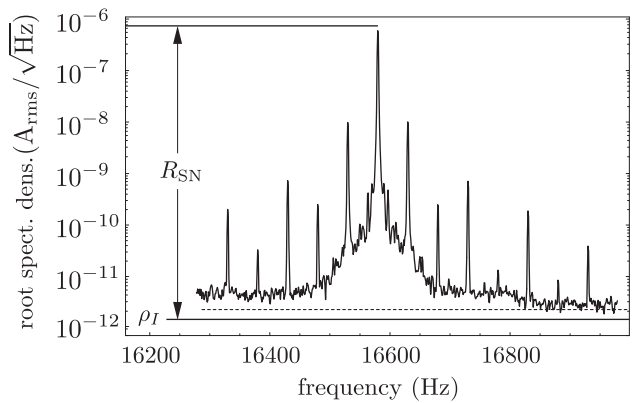}} \caption{
Root power spectrum of the photocurrent when the driving field is
in resonance with the Larmor frequency.
The data sample was recorded at 54\degree~C under conditions
optimized for maximum magnetometric sensitivity with a resolution
bandwidth of 1~\Hz{} (sampling time 1~s).
The amplitude measured by the lock-in amplifier corresponds to the
upper horizontal line.
The amplitude of the central peak is depressed, since it is
slightly broadened by the Hanning window used by the FFT spectrum
analyzer (see text).
The level $\rho_I$ is the shot-noise level calculated from the
DC-photocurrent.
The dashed line marks the rms noise measured at 23~k\Hz.
The \snr{} with respect to the calculated shot-noise level is $5
\times 10^5$.
The rms noise is a factor of 1.55 higher than $\rho_I$ resulting
in a \snr{} of $3.2\times 10^5$. } \label{fig:sn}
\end{figure}

An extended-cavity diode laser outside the shielded room was used as a
light source.
The laser frequency was actively stabilized to the $F=4\rightarrow
F=3$ transition of the Doppler broadened Cs $\mathrm {D}_1$ line
(894~nm) using DAVLL spectroscopy \cite{DAVLL} in an auxiliary
cell.
The light was delivered to the magnetometer sensor proper by a
multimode fiber (800~\mum{} core diameter).
After being collimated, the light was circularly polarized by a
combination of a polarizing beam-splitter and a multiple-order
quarter-wave plate.
The circularly polarized light then passed through a glass cell
containing the Cs vapor and a buffer gas to prevent the atoms from
being depolarized by wall collisions.
The cell could be heated to 65\degree~C using hot air which flowed
through silicon tubes wrapped around the cell holder.
The light power, $P$, transmitted through the glass cell was detected
by a photodiode specially selected to contain no magnetic materials.
A current amplifier (FEMTO Messtechnik, model DLPCA-200) converted the
photocurrent into a voltage that was fed to the input of the lock-in
amplifier.
The detection method resulted in a noise level 5~to 20\% above the
electron shot noise in the photodiode (Fig.~\ref{fig:sn}).
The digital lock-in amplifier (Stanford Research Systems, model
SR830) demodulated the oscillation of $P$ with reference to the
applied oscillating magnetic field.
That field was generated by two extra windings on each of the $B_x$
coils and was powered by the analog output of the reference
function generator contained within the lock-in amplifier.
The built-in function generator has the advantage that it delivers
a very pure sine wave (phase locked to the synchronization input)
and its amplitude can be controlled via the GPIB interface of the
lock-in amplifier.

In order to record magnetic resonance lineshapes the lock-in amplifier
was synchronized to a reference frequency supplied by a scanning
function generator.
The data measured by the lock-in (amplitudes of the in-phase and
quadrature signals) were transmitted in digital form to a PC, thus
avoiding additional noise.

\section{Optimization}\label{sec:optimization}

Although the theory of optical magnetometry is well known
\cite{bloom}, predictions about the real performance of a
magnetometer, especially when it is operating in weakly shielded
environments, are difficult to make.
The performance depends on laser power, rf power, cell size, laser
beam profile, buffer-gas pressure, and the temperature-dependent
density of Cs atoms.
The size of the cells and the buffer gas pressure were dictated by the
available cells: We used 20~mm long cells with 20~mm diameter
including 45~mbar Ne~and 8~mbar Ar~with a saturated Cs vapor.
Since the cell is oriented at 45\degree\ with respect to \Bstat,
the transverse spatial resolution was 28~mm.
The  cross section of the laser beam was limited by the 8~mm
apertures of the optical components (polarizers and quarter-wave
plates).

\subsection{Intrinsic resolution}

Our magnetometer produces a signal which was proportional to the
magnetic field changes.
The noise of the signal in a perfectly stable field therefore
determines the smallest measurable magnetic field change, called
the noise equivalent magnetic field (NEM).
The NEM is given by the square root, $\rho_B$, of the power spectral
density, $\rho_B^2$, of the magnetometer signal, expressed in
$\T/\Sqrt{\Hz}$.
The rms noise, $\sigma_B$, of the magnetometer in a given bandwidth
$\fbw$ is then
\begin{equation}
\sigma_B = \rho_B \sqrt{\fbw}.
\end{equation}
A straightforward way to measure the intrinsic sensitivity would
be to extract the noise level from a sampled magnetometer time
series via a Fourier transformation.
However, that process requires very good magnetic shielding since the
measured noise is the sum of the magnetic field noise and the
intrinsic noise of the magnetometer.
Many studies under well-shielded conditions have been carried out in
our laboratory, leading to the result that optical magnetometers are
in principle sensitive enough to measure the magnetic field of the
human heart.
However, the shielding cylinders used in these investigations were
too small to accommodate a person.
The present study investigates which level of performance can be
obtained in a weakly shielded environment with a volume large enough
to perform biomagnetic measurements on adults.
In the walk-in shielding chamber available in our laboratory the
magnetic noise level was about one order of magnitude larger than
the strongest magnetic field generated by the heart.
In order to compensate for this the actual cardiomagnetic
measurements were done with two magnetometers in a gradiometric
configuration \cite{bison1}.
However, the optimal working parameters where determined for a single
magnetometer channel only.

Since all time series recorded in this environment are dominated by
magnetic field noise, the straightforward way of measuring the
intrinsic noise could not be applied.
As an alternative approach a lower limit for the intrinsic noise can
be calculated using information theory.
The so-called \cramer{} lower bounds \cite{rife} gives a lower
limit on how precisely parameters, such as phase or frequency, can
be extracted from a signal in the presence of a certain noise
level.
For the following discussion we assume that the signal is a pure sine
wave affected by white noise with a power spectral density of
$\rho^2$.
We define the signal-to-noise ratio \snr{} as the rms amplitude, $A$,
of the sinusoidal signal divided by the noise amplitude, $\sigma$, for
the measurement bandwidth, $\fbw$:
\begin{equation}
\snr = \frac{A}{\sigma}=\frac{A}{\rho\sqrt{\fbw}}\: .
\label{eq:snr}
\end{equation}

For a magnetometer generating a Larmor frequency proportional to the
magnetic field, Eq.~(\ref{eq:gf}), the ultimate magnetic sensitivity
is limited by the frequency measurement process.
The \cramer{} lower bound for the variance, $V_\omega$, of the
frequency measurement \cite{rife} is used (Appendix~\ref{sec:CRf})
to calculate $\rho_B$
\begin{equation}
 \rho_B
 = \frac{\sigma_B}{\sqrt{ \fbw}}
 = \frac{4 \sqrt{3} \sqrt{\fbw}}  {\gamma_F \snr{}}\:.
\label{eq:rhof}
\end{equation}
For cardiac measurements a bandwidth of $\fbw = 100\:\Hz$ is required.
This together with a typical value for \snr{} of $10^4$ results in
a magnetic field resolution of $315\:\fT/\Sqrt{\Hz}$.
In order to be competitive with SQUID-based cardiomagnetometers that
feature an intrinsic noise of 5\ldots20~$\fT/\Sqrt{\Hz}$ this level of
performance is not sufficient.
For that reason we have concentrated on a different mode of operation
where the phase signal is measured by digital lock-in detection.

In this mode of operation $\omega\rf$ has a fixed value near the
Larmor frequency.
The information about the magnetic field is obtained from the phase
shift of the magnetometer response at that frequency.
The \cramer{} bound for a phase measurement of a signal with known
frequency is used in Appendix~\ref{sec:CRphi} to calculate the NEM for
that case:
\begin{equation}
\rho_B= \frac{ \Gamma_2 }{\gamma_F \snr\sqrt{\fbw}}\:.
\label{eq:rhophi}
\end{equation}
Equations~(\ref{eq:rhof}) and~(\ref{eq:rhophi}) define the bandwidth:
\begin{equation}
f_{0}=  \frac{\Gamma_2}{4\sqrt{3}}\:,
\end{equation}
for which both approaches yield the same magnetometric sensitivity.
For bandwidths larger than $f_0$, a phase measurement is more
advantageous whereas for bandwidths smaller than $f_0$ a frequency
measurement gives the higher sensitivity.

\subsection{Bandwidth}

\begin{figure}
\centerline{\includegraphics{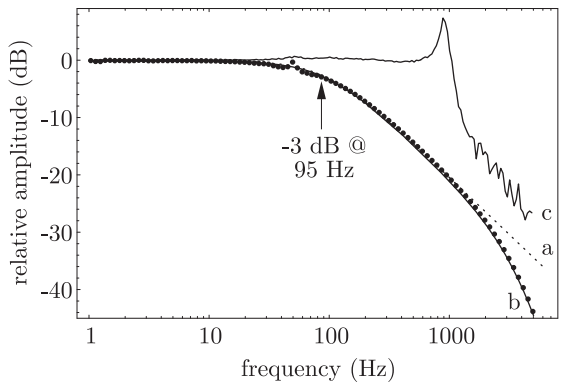}} \caption{
Frequency response of the magnetometer, measured by recording the
response to an oscillating magnetic field generated by a test
coil.
The dots show measured points recorded in free running mode under
conditions optimized for maximal magnetometric resolution.
(a) Calculated first-order low-pass filter corresponding to a spin
polarization lifetime of 1.67~ms.
(b) Fitted frequency response taking into account (a) and the
$4^\mathrm{th}$ order low-pass filter of the lock-in amplifier
(time constant $= 30~\mus$).
(c) Measured frequency response in the phase-stabilized mode. }
\label{fig:bw}
\end{figure}

In addition to the sensitivity, the bandwidth, i.e., the speed with
which the magnetometer signal follows magnetic field changes, is an
important feature of a magnetometer.
The steady-state solutions of the Bloch equations, \pqu\ and \pip\
[Eqs.~(\ref{eq:pip}) and~(\ref{eq:pqu})], follow small field changes
at a characteristic rate~$\Gamma_2$, corresponding to a delay time
$tau_S = \Gamma_2^{-1}$.
Since the steady state is only reached exponentially, the frequency
response is that of a first order low-pass filter [see
Fig.~\ref{fig:bw}(a)] with a (-3~dB) cut-off frequency $f_C$ given by
\begin{equation}
f_C = \frac{1}{2 \pi \tau_S} =\frac{\Gamma_2}{2 \pi} = \Delta \nu_2\:,
\end{equation}
and hence a bandwidth of
\begin{equation}
\fbw =   \frac{1}{4 \tau_S} = \frac{\Gamma_2}{4} = \frac{\pi}{2}\,
\Delta \nu_2 \:,
\label{eq:bw}
\end{equation}
where $\Delta \nu_2$ is the half width of the phase signal measured
in~\Hz.

To achieve maximum sensitivity, atomic magnetometers aim at a
maximum~$\tau_S$, at the cost of a reduced bandwidth of typically a
few tenths of~\Hz.
A large bandwidth can be obtained by increasing the light power since
that leads to shorter $\tau_S$ and therefore to higher bandwidth.
Larger light powers also increase the S/N ratio but the effect can be
overcompensated by magnetic resonance broadening, resulting in a
degradation of the magnetometric resolution.

Using feedback to stabilize the magnetic resonance conditions is
another way to increase the bandwidth.
Figure~\ref{fig:bw}(c) shows the frequency response of the OPM in both
the free-running (without feedback) mode and in the phase-stabilized
mode where the phase signal is used to stabilize $\omega\rf$ to the
Larmor frequency~$\omega_L$.
For large loop gain the bandwidth is mainly limited by loop delays.

A third method to achieve large bandwidths is the so-called
self-oscillating mode.
In this mode the oscillating signal measured by the photodiode is not
demodulated but rather phase-shifted and fed back to the rf-coils.
For a 90\degree\ phase shift the system then oscillates at the Larmor
frequency.
In order to measure the magnetic field, the frequency of this
oscillation has to be measured.
Magnetic field changes then show up --- at least theoretically
 \cite{bloom} --- as instantaneous frequency changes.
In practice, reaction times smaller than a single Larmor period
have been observed \cite{Dyal}.

Of the three modes outlined above, the latter two both rely on
frequency measurements.
The self-oscillating magnetometer provides a frequency that has to be
measured.
The phase-stabilized magnetometer measures the frequency via a
reference frequency locked to the Larmor frequency.
As a consequence, both methods suffer from the reduced magnetometric
resolution predicted by Eq.~(\ref{eq:rhof}).
Therefore, we have concentrated on the free-running mode of operation
for which the magnetometric resolution is given by
Eq.~(\ref{eq:rhophi}) and the bandwidth by Eq.~(\ref{eq:bw}).

Thanks to the rather high light power required for optimal
magnetometric resolution at higher cell temperatures, the cut-off
frequency of the magnetometer was 95~\Hz.
The bandwidth of the device under these conditions can be extracted
from the transfer function (Fig.~\ref{fig:bw}(b)) and is about 140~\Hz.
Because of the time constant of the lock-in amplifier, the measured
bandwidth is 10~\Hz smaller than the $\pi/2 \times 95 \:\Hz$ one
would expect for a first order low-pass filter [Eq.~(\ref{eq:bw})].

\subsection{Experimental lineshapes}
\label{sec:circle}

\begin{figure}
\centerline{\includegraphics{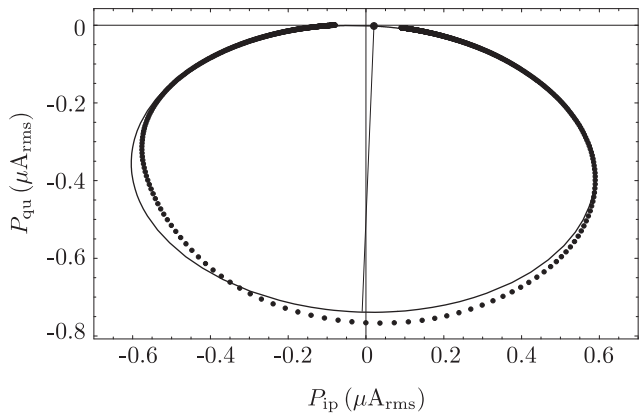}} \caption{
The dots represent a Nyquist plot of the magnetic resonance
measured under optimized conditions.
The solid line is a fit of Eq.~(\ref{eq:tgrad}) with added offset
and phase rotation to the measured data.
The fit model assumes a constant magnetic field distribution.
The offset is indicated by the dot close to the origin.
The short diameter of the ellipse is drawn in order to illustrate
the phase rotation of 2.4\degree.
 } \label{fig:circ}
\end{figure}

Figure \ref{fig:circ} shows a Nyquist plot with experimental data and
a model simultaneously fit to the in-phase and quadrature components
of the data.
The data show a certain asymmetry that can not be reproduced by the
model.
The Nyquist plots for different magnetic field distributions
(Fig.~\ref{fig:circ_grads}) suggest that the asymmetry is caused by
inhomogeneous magnetic fields.
Unfortunately the models discussed in Sec.~\ref{sec:nyquist} do not
fit the data correctly, implying that higher-order gradients cause the
deformation of the measured lineshape.
The fact that the asymmetry is more pronounced for high rf amplitudes
indicates that inhomogeneous rf-fields --- causing the different parts
of the ensemble to contribute with different widths --- have to be
considered.
Unfortunately, models for such inhomogeneities do not lead to analytic
line shapes.
An empirical model which assumes the measured resonance consists of a
sum of several resonances, each at a different position and with a
different width, can be fit to the data.
The data can be fit perfectly if the number of subresonances is high
enough.
However, such fits have a slow convergence and do not provide the
needed information about the width and amplitude of the resonance in
single fit parameters.
For practical reasons (during the optimization more than 2000 spectra
were fit) we decided to use the constant magnetic field distribution
model for fitting data similar to the ones in Fig.~\ref{fig:circ}.

Magnetic field inhomogeneities have much less influence on the shape
of the phase signal resulting in more reliable values for $\Gamma_2$.
The phase signal represents the speed with which the resonance evolves
through the Nyquist plot.
Using both the phase signal and the Nyquist plot, the in-phase and
quadrature components of the resonance were reconstructed, however,
the frequency scaling were given by the phase signal only.

\subsection{Optimization measurements}
\label{sec:optmes}

For the optimization of the NEM given by Eq.~(\ref{eq:rhophi}) the S/N
ratio of the lock-in input signal and the linewidth~$\Gamma_2$ have to
be measured.
Figure~\ref{fig:sn} shows a frequency spectrum recorded at the input
of the lock-in amplifier using a FFT spectrum analyzer.
The frequency $\omega\rf$ was tuned to the center of the magnetic
resonance so that the modulation of the photocurrent was at its
maximum amplitude.
The power spectrum shows a narrow peak at $\omega\rf$ surrounded by
noise peaks that characterize the magnetic field noise.
Monochromatic magnetic field fluctuations, e.g., line frequency
interference, modulate the phase of the measured sine wave and show up
in the power spectrum as sidebands.
The low frequency flicker noise of the magnetic field thus generates a
continuum of sidebands that sum up to the background structure
surrounding the peak in Fig.~\ref{fig:sn}.
The estimation of the intrinsic sensitivity is based on the assumption
that those sidebands would disappear in a perfectly constant magnetic
field.
The amplitude noise of the signal is mainly due to the electron shot
noise in the photodiode, which generates a white noise spectrum.
For frequencies which are more than 1~k\Hz{} away from the resonance,
the noise level drops to the white noise floor.
The electron shot noise is the fundamental noise level that can not be
avoided.
The noise spectral density $\rho_I$ can be calculated from the DC
current~$I_\mathrm{DC}$ flowing through the photodiode:
\begin{equation}
\rho_I =  \sqrt{2 e I_\mathrm{DC}}\:.
\end{equation}

At room-temperature the measured rms noise in the spectrum was 5\% to
20\% above the shot-noise level, depending on induced noise on the
photocurrent and the laser frequency stabilization that could cause
excess noise in the light intensity.
The rms noise rose rapidly for higher temperatures because of the
increasing leakage current in the photodiode.
Unfortunately, in the experimental setup the photodiodes were in good
thermal contact with the Cs cell and, given that the optimal operating
temperature of the Cs cells turned out to be in the range of
50\degree~C to 60\degree~C, the photodiode produced an excess noise
larger than the shot noise of the photocurrent.
Figure~\ref{fig:sn} shows a spectrum recorded under conditions
optimized for maximal magnetometric resolution.
At 53\degree~C the measured rms noise was higher than the shot noise
by a factor of 1.55.
However, this limitation can be overcome easily since the photodiodes
do not need to be close to the Cs cell and thus can be operated at
room-temperature.
In order to avoid the problem of drifting values of~$\rho$ during the
optimization of $\rho_B$, the theoretical shot noise level~$\rho_I$
was used for~$\rho$ in Eq.~\ref{eq:snr} instead of the measured noise.

The amplitude~$A$ of the signal can be extracted from the FFT-spectrum
by integrating the spectrum over three points ($\pm 1\:\Hz$) around
the center frequency.
The procedure was needed since the Hanning window used by the spectrum
analyzer to reconstruct the spectrum causes a slight broadening of the
central peak.
The values calculated in that way are in good agreement with those
measured by the lock-in amplifier.

The third parameter needed to calculate the intrinsic sensitivity is
the half-width~$\Gamma_2$ of the magnetic resonance.
The value was extracted from a magnetic resonance spectrum recorded by
the lock-in amplifier during a frequency sweep of the applied
oscillating magnetic field.
As discussed in Section~\ref{sec:circle} a constant-gradient model was
fit to the data in order to extract $\Gamma_2$.

For optimizing in a three-dimensional parameter space, the time for
one measurement had to be kept as short as possible.
When the lock-in amplifier signal was used as a measure for $A$ [see
Eq.~(\ref{eq:snr})] and the noise was calculated from the DC current
it was not necessary to record a FFT spectrum for each set of
parameters of the optimization procedure.
In that way the time for a single NEM measurement was reduced to the
20~s sweep time of $\omega\rf$ plus the time needed to measure the DC
current and the temperature of the cell.
The measurement was controlled by a PC running dedicated software for
recording and fitting the magnetic resonance signals.
The amplitude of the rf field, $B\rf$, was changed automatically by
the software, resulting in series of typically ten NEMs as a function
of $B\rf$.
A typical optimization run was made by recording many such series
while the system slowly heated up.
Repeating those runs for different light powers finally resulted
in data for the whole parameter space.

\section{Results}

\subsection{Dependence on rf amplitude}

\begin{figure}
\centerline{\includegraphics{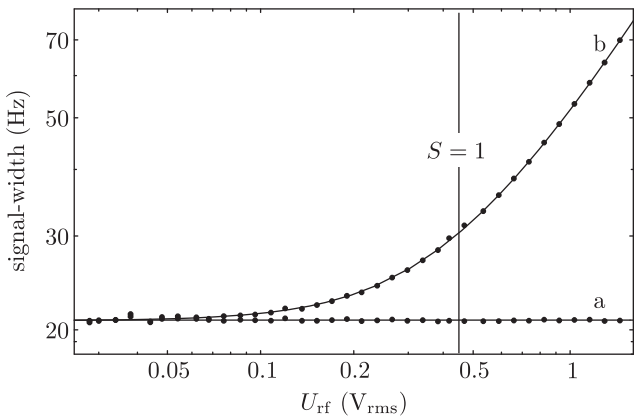}} \caption{
Dependence of the magnetic resonance linewidth on the rf amplitude
$U\rf$.
The points are extracted from measured magnetic resonance spectra
by least squares fitting of model Eq.~(\ref{eq:tgrad}).
The phase signal (a) has a constant linewidth whereas the common
widths of the in-phase and quadrature signals (b) increase rapidly
with rf amplitude.
The solid line represents a model fitted to the data that assumed
an additional broadening caused by inhomogeneous magnetic fields.
 } \label{fig:xywidth}
\end{figure}

The first study made with the magnetometer examined the dependence of
the magnetic resonance on the rf amplitude $B\rf$ measured a series of
spectra recorded at room temperature.
Figure~\ref{fig:xywidth} shows the dependence of the magnetic
resonance signal width on the rf amplitude measured by the coil
voltage $U\rf$.
The width of the phase signal (see Fig.~\ref{fig:xywidth}) was fit
with a constant, whereas the common width of the in-phase and
quadrature components were given by Eq.~(\ref{eq:deltaomega}).
To fit the widths correctly, a constant width had to be added to
Eq.~(\ref{eq:deltaomega}).
The additional constant width can be interpreted as a residual
broadening caused by magnetic field inhomogeneities of higher order
than the one considered in the line fitting model.
The Nyquist plot (see Fig.~\ref{fig:circ}) shows that higher order
gradients are present and the excellent agreement in
Fig.~\ref{fig:xywidth} suggests that they can be treated as an
additional broadening.

\begin{figure}
\centerline{\includegraphics{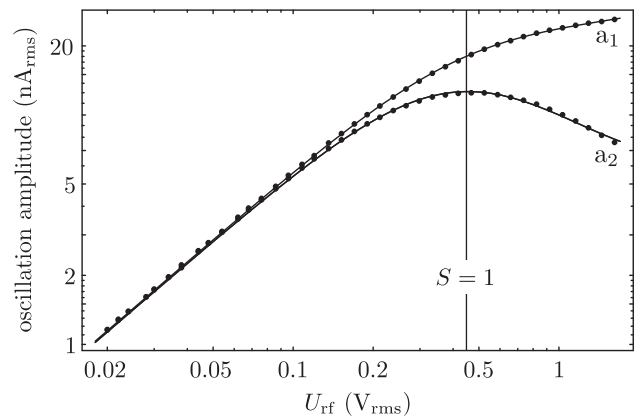}} \caption{
Amplitude of the in-phase ($a_1$) and quadrature~($a_2$) signals
as a function of rf amplitude $U\rf$.
The points represent values extracted from measured magnetic
resonance spectra.
The solid lines show a model fit to the data points (see text).
The quadrature amplitude $a2$ is equal to the amplitude of the
incoming sine wave on resonance ($\delta = 0$). }
\label{fig:xyampl}
\end{figure}

Figure~\ref{fig:xyampl} shows the amplitudes of the in-phase and
quadrature magnetic resonance signals.
The amplitudes where extracted from the same spectra used for
Fig.~\ref{fig:xywidth}.
The fit model used to explain the amplitudes (solid lines in
Fig.~\ref{fig:xyampl}) was based on Eqs.~(\ref{eq:dr})
and~(\ref{eq:di}) with a background proportional to $B\rf$.
The origin of the background was an inductive pick up of the \Bosci{}
field by the photocurrent loop which caused an additional
phase-shifted sine wave to be superposed on the photocurrent.
As discussed in the theory part (see Fig.~\ref{fig:circ_phase}) that
lead to an offset in the measured amplitudes of the magnetic resonance
signal.

The NEM as a function of rf amplitude is inversely proportional to the
quadrature amplitude ($a_2$ in Fig.~\ref{fig:xyampl}), since the
linewidth of the phase signal and the shot noise do not change with rf
amplitude.
The optimal rf amplitude was determined from the data shown in
Fig.~\ref{fig:xyampl} and corresponds to $S=1$.

\subsection{Dependence on temperature and light power}

\begin{figure}
\centerline{\includegraphics{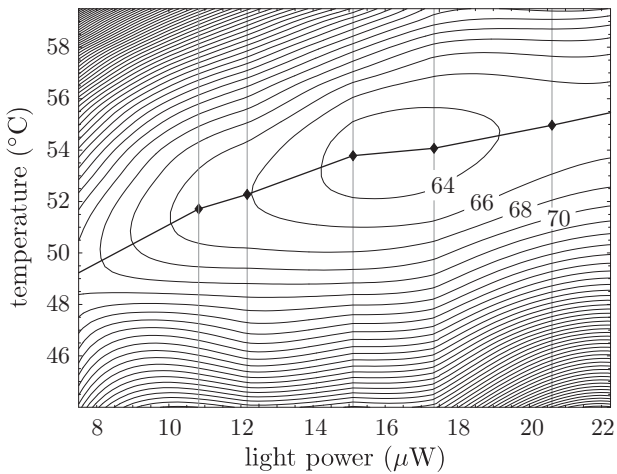}} \caption{
Contour plot of the magnetometric resolution (NEM) as a function
of temperature and light power.
The map is calculated by numeric interpolation from seven
optimization runs (indicated by vertical lines).
The labels at the contours mark the NEM in $\fT/\Sqrt{\Hz}$.
The points of minimal NEM for each optimization run are indicated
by points.
The connecting line is a cut along which the data of
Fig.~\ref{fig:4plots} are obtained.
Including the variation of the rf amplitude 970 parameter sets
were recorded and analyzed to produce the map.
 } \label{fig:map}
\end{figure}

\begin{figure}
\centerline{\includegraphics{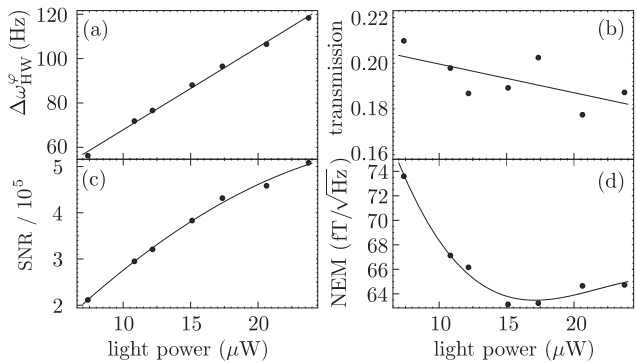}} \caption{
Magnetic resonance parameters as a function of incident light
power:
(a)~shows the width of the phase signal that determines the
cut-off frequency of the magnetometer bandwidth;
(b)~shows the DC transmission through the Cs cell relative to the
incident light power;
(c)~is the signal/noise ratio of the lock-in input signal with
respect to calculated shot noise;
(d)~shows the NEM at the points indicated in Fig.~\ref{fig:map}. }
\label{fig:4plots}
\end{figure}

As described in section \ref{sec:optmes} the dependence of the NEM on
the temperature was recorded while the system was slowly heated.
The rf amplitude was automatically scanned so that for every
temperature the optimal rf amplidude could be determined.
Figure~\ref{fig:map} shows a contour plot of the NEM as a function of
temperature and light power.
If the light power is increased, the temperature (and hence Cs atom
density) has also to be increased to maintain optimal resonance
conditions.
Figure~\ref{fig:4plots}(b) shows the power transmitted through the
cell relative to the incident light power.
A relative transmission of 0.37 corresponds to an absorption length
which matches the cell length.
Taking into account losses at the windows, a density corresponding to
1.4 absorption lengths was found to be optimal.

For each light power the optimal temperature is indicated by a dot in
Fig.~\ref{fig:map}.
Plotting the NEM along the optimum temperature power curve, i.e., the
curve connecting the dots, results in the plot shown in
Fig.~\ref{fig:4plots}(d).
For light powers below 15~$\mu$W and the corresponding
temperatures the sensitivity rapidly degrades.
The loss in sensitivity is less pronounced if the power and
temperature are chosen above the optimum.
Values for \snr\ of up to 500~000 (114~dB) were measured at a
resolution bandwidth of 1~\Hz.

The optimal magnetic field resolution of our magnetometer is reached
at a light power of 15~$\mu$W and a temperature of 53\degree~C.
With that set of parameters, the usable bandwidth of the magnetometer
was determined by a cut-off frequency of about 80~\Hz{} (see
Fig.~\ref{fig:4plots}(a)).
In order to meet the required 100~\Hz{} bandwidth a slightly larger
light power can be used.
All characterizing measurements (cf. Figs.~\ref{fig:res}, \ref{fig:bw},
\ref{fig:circ}, and \ref{fig:sn}) were therefore performed with a
light power of 20~$\mu$W at 54\degree~C.

\section{Conclusion}

Optimizing the performance of the magnetometer has led to a set of
parameters for which the device offers a large sensitivity and a large
bandwidth.
Both requirements can be met at the same time because of rather large
linewidths that turned out to be optimal.
Under these conditions the high magnetometric sensitivity relies on
the achieved very high signal/noise ratios.
The system has the potential to operate at a \snr\ of 500000
(Fig.~\ref{fig:sn}) and we hope to be able to demonstrate this once
the photodiodes can be removed from the heated Cs cell.
However, even using the measured \snr{} of 320000, the intrinsic
sensitivity of $100\:\fT/\Sqrt{\Hz}$ is good enough for less demanding
cardiomagnetic measurements.

The magnetometer bandwidth of 140~\Hz{} in the free-running
phase-detecting mode (Fig.~\ref{fig:bw}) is high enough for cardiac
measurements.
The phase-detecting mode avoids the fundamental limitations
associated with frequency measurements using short integration
times.

An important open experimental question is whether the predicted
intrinsic sensitivity can be reached using several of the present OPMs
in a higher order gradiometer geometry.
With gradiometric SQUID sensors it is possible to achieve NEM
value on the order of $20\:\fT/\Sqrt{\Hz}$ in unshielded
environments \cite{fenici1}.
In future we plan to use cells with spin-preserving wall coatings
rather than buffer-gas cells as sensing elements.
Coated cells have the advantage that the atoms traverse the volume
many times during the spin coherence lifetime, therefore averaging
out field inhomogeneities.
We are therefore confident that the present limit from field gradients
can be overcome and that optical magnetometers can reach an operation
mode limited by their intrinsic sensitivity.


\section{Appendix A: The \cramer{} bound for frequency measuring magnetometers}
\label{sec:CRf}

For the measurement of the frequency $\omega$ of a sine wave with
a rms amplitude $A$ sampled at $N \gg 1$ points separated by time
intervals~$T_s$ the \cramer{} lower bound for the
variance~$V_\omega$ of $\omega$ in the presence of white Gaussian
amplitude noise of variance~$\sigma^2$ is given by \cite{rife}:
\begin{equation}
V_\omega = \frac{12 \sigma^2}{A^2 T_M^2 N},
\end{equation}
where $T_M=N T_s$ is the total time interval for one frequency
determination.
The bandwidth on the input side of the lock-in amplifier is therefore
$\Fbw=1/2T_s=N/2 T_M$, that at the output is $\fbw=1/2 T_M$.
With the definition of the signal-to-noise ratio, Eq.~(\ref{eq:snr}),
$V_\omega$ can be expressed independently of the number of samples:
\begin{equation}
V_\omega = \frac{12\rho_i^2\Fbw}{A^2T_M^2N}=\frac{6}{\snr^2 T_M^3}.
\end{equation}
Ideal measuring processes are limited by that condition only.
Frequency measurements by a FFT with peak interpolation is a
\cramer{} bound limited measuring process \cite{rife}.

From that bound a lower limit for the performance of a frequency
measuring magnetometer can be derived.
The so-called self-oscillating $M_x$ magnetometer \cite{bloom} is
of this type since it supplies an oscillating signal with a
frequency proportional to the magnetic field.
With Eq.~(\ref{eq:gf}) it follows that the root spectral density of
the measurement noise $\rho_B$ is given by:
\begin{equation}
 \rho_B
 = \sqrt{\frac{V_B}{ \fbw}}
 = \frac{4 \sqrt{3} \sqrt{\fbw}}  {\gamma_F \snr{}}.
\end{equation}

\section{Appendix B: The \cramer{} bound for phase measuring magnetometers}
\label{sec:CRphi}

The \cramer{} lower bound for the measurement of the phase of a
signal with known frequency is given by \cite{rife}:
\begin{equation}
V_\varphi = \frac{\sigma^2}{A^2 N}\:.
\end{equation}
An example of a measurement process limited only by that condition is
the lock-in phase detection where the phase is calculated from the
in-phase and quadrature outputs of the lock-in amplifier [see
Eq.~(\ref{eq:phase})].
In order to calculate the variance $V_\varphi$ of the phase
measurement we assume a white amplitude noise spectrum with a power
spectral density $\rho^2$:
\begin{equation}
V_\varphi = \frac{\rho^2}{A^2 2 T_M}\: .
\label{eq:vphi}
\end{equation}
Using this expression and the definition of \snr{}
[Eq.~(\ref{eq:snr})], Eq.~(\ref{eq:vphi}) can be written as
\begin{equation}
\sigma_\varphi^2 = V_\varphi
= \frac{\rho^2 \fbw}{A^2}
= \frac{\fbw}{\snr^2}\:.
\label{eq:sigmaphi}
\end{equation}
From the measured phase $\varphi$, the detuning
$\delta=\omega\rf-\omega_{L}$ can be derived.
For $\varphi \ll 1$, Eq.~(\ref{eq:phase}) leads to $\delta\approx
\Gamma_2\varphi$.
Using Eq.~(\ref{eq:gf}), the detuning can be expressed as a magnetic
field difference $\Delta B =\delta/\gamma$ which leads, together with
Eq.~(\ref{eq:sigmaphi}), to the magnetic field resolution $\sigma_B$:
\begin{equation}
 \sigma_B
 = \frac{\sigma_\delta}{\gamma_F}
 = \frac{\sigma_\varphi \Gamma }{\gamma_F }
 = \frac{\Gamma \sqrt{\fbw}}{\gamma_F \snr\sqrt{\fbw}}\:.
\end{equation}
The root spectral density of the noise in the $\Delta_B$ measurement,
$\rho_B = \sigma_B/ \Sqrt{\fbw}$, is thus given by:
\begin{equation}
\rho_B= \frac{ \Gamma }{\gamma_F \snr\sqrt{\fbw}}\:.
\end{equation}

\section*{Acknowledgments}
This work was supported by grants from the Schweizerischer
Nationalfonds and the Deutsche Forschungsgemeinschaft.
The authors wish to thank Martin Rebetez for efficient help in
understanding the frequency response of the magnetometer and Paul
Knowles for useful discussions and a critical reading of the
manuscript.


\end{document}